\documentclass[11pt]{article}
\usepackage{hyperref}
\usepackage{graphicx}
\pdfoutput=1
\begin{document}
\title{A Numerical Simulation of a Plunging Breaking Wave}
\author{Paul Adams$^1$, Kevin George$^1$, Mike Stephens$^1$,\\
Kyle A. Brucker$^2$, Thomas O'Shea$^2$ and Douglas Dommermuth$^2$\\ 
\vspace{4pt} \\
$^1$ Unclassified Data Analysis and Assessment Center, U.S. Army\\
Engineering Research and Development Center, MS 39180 \\
\vspace{2pt}\\
 $^2$  Science Applications International Corporation\\
10260 Campus Point Drive, San Diego, CA 92121}
\date{November, 22 2009}
\maketitle

\begin{abstract}
  This article describes the fluid dynamics video, ``A Numerical Simulation of a Plunging Breaking Wave'', which was submitted to the gallery of fluid motion at the 2009 APS/DFD conference.  The simulation was of a deep-water plunging breaking wave.  It was a two-phase calculation which used a Volume of Fluid (VOF) method to simulate the interface between the two immiscible fluids.  Surface tension and viscous effects were not considered.  The initial wave was generated by applying a spatio-temporal pressure forcing on the free surface.  The video shows the 50\% isocontour of the volume fraction from several different perspectives.  Significant air entrainment is observed as well as the presence of stream-wise vortex structures.
\end{abstract}

\section{Formulation}
\begin{figure}
\begin{center}
\includegraphics[width=8cm]{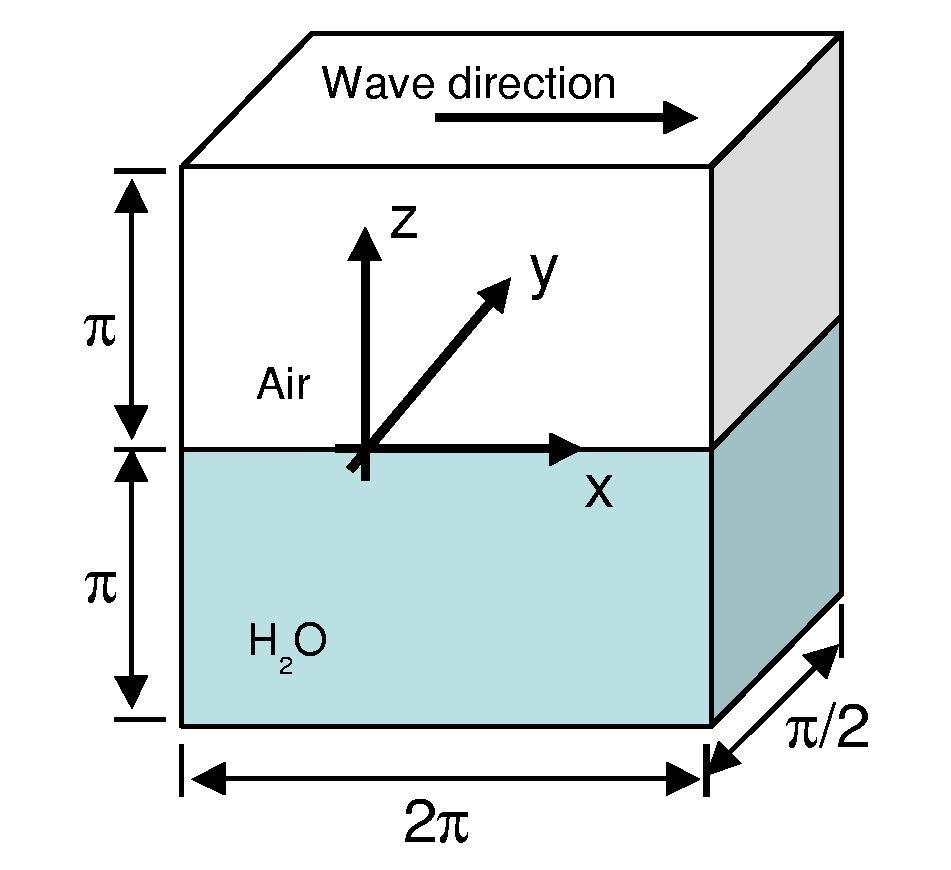}
\end{center}
\caption{Schematic of computational domain. $\vec{g}=[0,0,-g]$.}
\label{fig:setup}
\end{figure}

 The computational domain, shown schematically in Fig.~\ref{fig:setup}, moves with the linear crest speed of the wave, $U$, and has dimensions $[2\pi,\pi/2,2\pi]$, with $[1024,256,512]$ grid points. Periodic boundary conditions are used in the horizontal directions and free-slip boundary conditions in the vertical.  The Froude number, $Fr=U(k/g)^{1/2}$, is unity, where $g$ is the acceleration due to gravity and $k$ is the wave number.  The density ratio between the two fluids is 1000:1.  
  The total distance traveled by the wave, $X$ in the laboratory frame, is equivalent to $U t$ in the simulation.  Gravity acts in the $-z$ direction.

\section{Numerical Method}
Let $u_i^*$ denote the three-dimensional velocity field, $\rho$ the density, and $p$ the pressure, all functions of space ($x_i^*$) and time ($t^*$), where an asterisk denotes a dimensional quantity.  Time, space, velocity, density and pressure are normalized as follows:
\begin{equation}
t= t^{*} U k \quad,\quad x_{i} =  x_{i}^{*}k \quad,\quad u_{i}=\frac{u_{i}^{*}}{U} \quad,\quad \rho=\frac{\rho^{*}}{ \rho_w } \quad, \quad p = \frac{{p}^{*}}{ \rho_w U^2} \, .
\label{nondim}
\end{equation}
where $\rho_w$ is the density of water, $k$ is the wave number, and $U=\left(g/k\right)^{1/2}$.

  In the Volume of Fluid method (Rider {\it et al.} (1994)) the fraction of fluid that is inside a cell is denoted by $\phi$.  By definition, $\phi=0$ for a cell that is totally filled with air, and $\phi=1$ for a cell that is totally filled with water.  The density expressed in terms of $\phi$ is
\begin{equation}
\rho = (1-\lambda)\phi+\phi 
\label{eq:rho}
\end{equation}
where $\lambda$ is the density ratio between air and water.

\vspace{12pt}
\noindent The non-dimensional governing equations are:\\
{\bf Momentum:}
\begin{equation}
\frac{\partial u_{i}}{\partial t} + \frac{\partial \left(u_{k}u_{i}\right)}{\partial x_{k}} = -\frac{1}{\rho} \frac{\partial p}{\partial x_{i}} -\frac{1}{Fr^2} \delta_{i3} + \frac{p_a}{\rho}\delta\left(\phi\right),
\label{eq:mom}
\end{equation}
{\bf VOF:}
\begin{equation}
\frac{\partial \phi}{\partial t} + u_{k}\frac{\partial \phi}{\partial x_{k}} = 0.
\label{eq:vof}
\end{equation}
 The relevant non-dimensional parameter is the surface Froude number, $Fr=U(k/g)^{1/2}$.  The temporal integration is handled with an explicit RK2 scheme, and the advective terms with the flux based limited QUICK scheme of Leonard (1997).  The VOF algorithm uses the operator-split method of Puckett {\it et al.} (1997).  As discussed in Dommermuth {\it et al.} (1998) the divergence of the momentum equation, Eq.~(\ref{eq:mom}), combined with the solenoidal constraint $\partial u_i/\partial x_i = 0$, provide a Poisson equation for the dynamic pressure. 

  The last term on the $r.h.s.$ of Eq.~(\ref{eq:mom}) is the atmospheric pressure forcing term.  In the simulation discussed here the following forcing function was used for $t<4\pi$
\begin{equation}
p_a = \frac{1}{2}\left[A_0cos\left(x\right)+A_r\right]\left[1-cos\left(\frac{t}{2}\right)\right]
\label{eq:pa}
\end{equation}
where
\begin{equation}
 0.1A_0=\left[\frac{1}{V}\int_V dVA_r\left(x_i\right)\right]^2
\label{eq:A0}
\end{equation}
Here, $A_0=0.02$ and $A_r$ is a uniform random disturbance that has been passed through a low-pass filter.

\section{Flow description}
\begin{figure}
\begin{center}
\includegraphics[width=8cm]{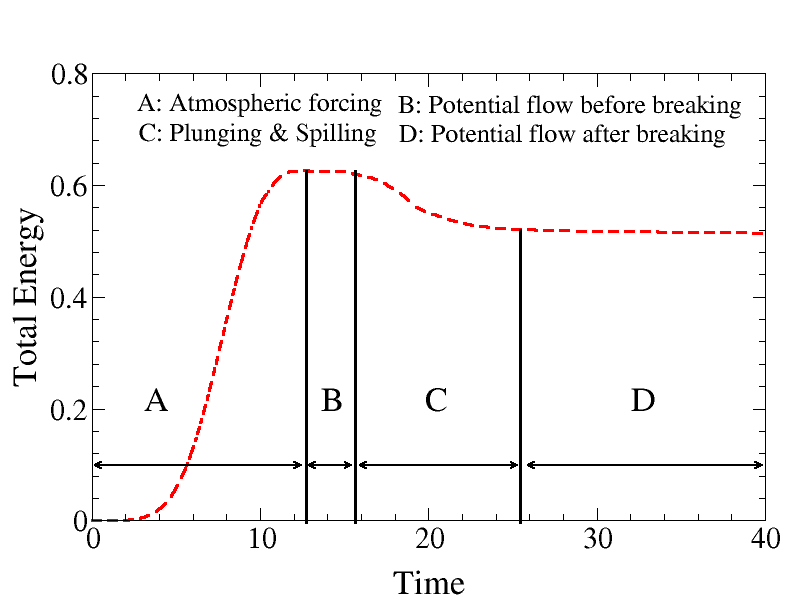}
\end{center}
\caption{Total Energy, $\int_V dV \rho \left(U_iU_i + g z\right)$ }
\label{fig:evolution}
\end{figure}

Fig.~\ref{fig:evolution} is a plot the the total energy, 
\begin{equation}
E(t)=\int_V dV \rho(t) \left(U_i(t)U_i(t) + g z \right),
\end{equation}
over time.  Four distinct stages are evident in Fig.~\ref{fig:evolution}, they are:  A. Atmospheric forcing; B. Potential flow before breaking; C. Breaking which consists of plunging, spilling and splash-up events; and D. Potential flow after breaking.  

\section{Video}
  The full size video, mpeg2 encoded, is approximately $75Mb$ (\href{http://ecommons.library.cornell.edu/bitstream/1813/13802/2/APS_Contest_1024x576_v3.mpg}{download}).  The web size video, mpeg1 encoded is approximately $10Mb$ (\href{http://ecommons.library.cornell.edu/bitstream/1813/13802/3/APS_Contest_768x432_short.m1v}{download}).  The videos are also available at \href{http://www.saic.com/nfa}{www.saic.com/nfa}.

Scenes 1 and 2 in the video start near the end of stage A (see Fig.~\ref{fig:evolution}) and show the evolution through stages B and C and end early in stage D.  Scene 3 starts after the initial plunging event, early in stage C and ends late in stage C.  Scene 4 starts near the end of stage B, and ends in the middle of stage C.  Scene 5 (Vortex Tubes) starts near the end of stage B and pauses/ends early in stage C.  In scenes 1, 2 and 3 the 50\% isocontour of the volume is shown.  In scenes 4 and 5 the magnitude of velocity projected onto the free surface is shown in selected regions.

\begin{figure}
\begin{center}
\includegraphics[width=8cm]{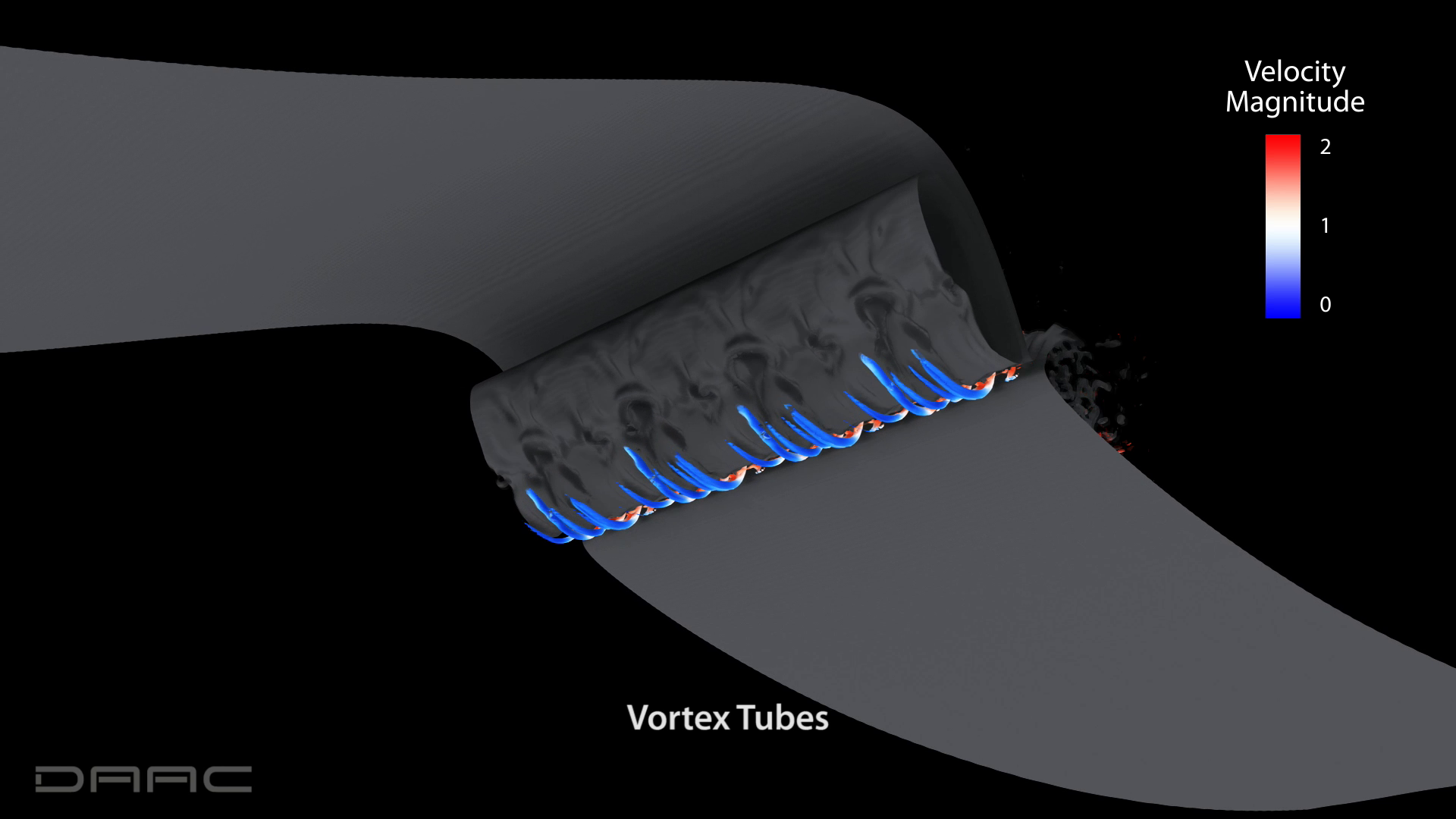}
\end{center}
\caption{Vortex tubes on the underside of the initial oval of air entrained by the first plunging event.}
\label{fig:still}
\end{figure}

\vspace{12pt}
\noindent{\bf SCENES:}
\begin{enumerate}
\item Tank View-- A perspective view with walls added to provide context to the breaking events.  After the initial plunging event, a jet is formed (splash-up).  After the secondary jet impinges on the free surface the remaining portion of stage C is a sequence of weaker and weaker plunging and spilling events, until the flow returns to a potential flow.\\
\item 4-way View-- Here the evolution is viewed from four angles simultaneously.  Note the large pocket of air that is entrained (lower right view), and the irregularity on the front face of the wave (lower left view).  \\
\item Side (bottom) Zoom-- This scene is a magnified view of the lower right frame from Scene 2, and shows the formation of the jet during splash-up.  Significant air entrainment is observed shortly after the impact of the splash-up jet.  This air entrained by the splash-up jet persists for a longer time than the initial ovular pocket of air entrained by the initial plunging event.\\
\item Velocity Magnitude-- This is Scene 4 with the magnitude of the velocity projected onto the free surface. 
\item Vortex Tubes-- Scene 4 paused to show the vortex tubes, and hairpin like structures on the back of the initial oval of air entrained by the plunging event.  Shown here as Fig.~\ref{fig:still}
\end{enumerate}

\section{Acknowledgments}
We would like to acknowledge Dr. Pat Purtell with the United States Office of Naval Research for the support through ONR grant N00014-07-C-0184.  This work was supported in part by a grant of computer time from the \href{http://www.hpcmo.hpc.mil}{DOD High Performance Computing Modernization Program}. The numerical simulations have been performed on the Cray XT3 and XT4 at the U.S. Army Engineering Research and Development Center.

\section{References}
Dommermuth, D.G., O’Shea, T.T., Wyatt, D.C., Ratcliffe, T., Weymouth, G.D., Hendrickson, K.L., Yue, D.K.P., Sussman, M., Adams, P., \& Valenciano, M. (2007) An application of cartesian-grid and volume-of-fluid methods to numerical ship hydrodynamics.   In the proc. of the 9th Int. Conf. on Num. Ship Hydro., Ann Arbor, MI, Aug. 5-8.\\

\noindent Dommermuth, D., Innis, G., Luth, T., Novikov, E., Schlageter, E., \& Talcott, J. (1998) Numerical Simulation of Bow Waves.  Proc. of the 22nd Symposium on Naval Hydrodynamics, Washington D.C., pp. 508-521.\\

\noindent Leonard, B., (1997) Bounded higher-order upwind multidimensional finite volume convection-diffusion algorithms.  W. Minkowycz \& E. Sparrow, eds.  Advances in Numerical Heat Transfer, Taylor and Francis, Washington D.C., pp. 1-57.\\

\noindent Rider, W., Kothe, D., Mosso, S. Cerutti, J. \& Hochstein, J. (1994) Accurate solution algorithms for incompressible multiphase flows.  AIAA paper 95-0699.\\

\noindent Puckett, E., Almgren, A., Bell, J., Marcus, D. \& Rider, W. (1997) A second-order projection method for tracking fluid interfaces in variable density incompressible flows.  J. Comp. Physics 130, pp. 269-282.\\
\end{document}